\documentclass{icrc2009}

\usepackage{graphicx}   
\usepackage[font=footnotesize,caption=false]{subfig}
\usepackage{fixltx2e}
\usepackage{url}

\newcommand{\shorttitle}[1]%
{\markboth{Proceedings of the 31\MakeLowercase{$^{st}$} ICRC, {\L}\'{o}d\'{z} 2009}{#1} }
\newcommand{\etal}{\MakeLowercase{\textit{et al. }}} 

\begin{document}
\title{A Search For Atmospheric Neutrino-Induced Cascades with IceCube}

\author{\IEEEauthorblockN{Michelangelo D'Agostino\IEEEauthorrefmark{1} for the IceCube Collaboration\IEEEauthorrefmark{2}}                           
                            \\
\IEEEauthorblockA{\IEEEauthorrefmark{1}Department of Physics, University of California, Berkeley, CA 94720, USA}
\IEEEauthorblockA{\IEEEauthorrefmark{2}See the special section of these proceedings.}}

\shorttitle{M. D'Agostino \etal Atmospheric Neutrino-Induced Cascades}
\maketitle

\begin{abstract}
The IceCube detector is an all-flavor neutrino telescope. For several years IceCube has been detecting muon tracks from charged-current muon neutrino interactions in ice. However, IceCube has yet to observe the electromagnetic or hadronic particle showers or ``cascades'' initiated by charged- or neutral-current neutrino interactions.  The first detection of such an event signature will likely come from the known flux of atmospheric electron and muon neutrinos. A search for atmospheric neutrino-induced cascades was performed using a full year of IceCube data.  Reconstruction and background rejection techniques were developed to reach, for the first time, an expected signal-to-background ratio $\sim$1 or better.  
 \end{abstract}

 
IceCube is a cubic kilometer neutrino telescope currently under construction at the geographical South Pole.  With 59 of 86 strings of photomultiplier tubes currently embedded into Antarctica's deep glacial ice, IceCube is already the world's largest neutrino detector~\cite{2008arXiv0812.3981K}.
 
IceCube detects high energy neturinos by observing Cherenkov light from the secondary particles produced in neutrino interactions in ice.  In charged-current $\nu_{\mu}$ interactions, the outgoing energetic muon emits light along its track through the detector.   A hadronic particle shower or cascade is also produced at the neutrino interaction vertex, but this is usually well outside of the instrumented detector volume.  In charged-current $\nu_{e}$ interactions, the outgoing electron initiates an electromagnetic (EM) cascade which accompanies the hadronic cascade.  Neutral-current interactions of any neutrino flavor produce hadronic cascades.  

At the energies relevant for atmospheric neutrinos, both hadronic and EM cascades develop over lengths of only a few meters.  In a sparsely instrumented detector like IceCube, they look like point sources of Cherenkov light whose spherical wavefronts expand out into the detector.  While muon tracks have been detected by neutrino telescopes, cascade detection has remained an elusive goal for high energy neutrino astrophysics. 

The well-studied atmospheric neutrino flux can serve as a calibration source for the cascade detection channel and should provide a valuable proof-of-principle for all-flavor detection.  Once neutrino-induced cascades have been detected from the atmosphere, they should also open up a powerful channel for astrophysics analysis.  Since cascades are topologically distinct from muons, they can be separated from the cosmic ray background over the entire $4\pi$ of the sky~\cite{Ignacio-GRB-cascades}.

  \begin{figure}[!t]
  \centering
  \includegraphics[width=2.5in]{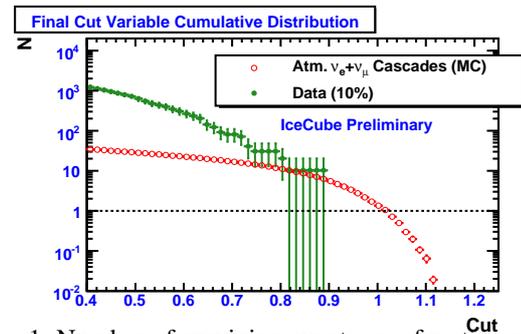}
   \caption{Number of surviving events as a function of final cut strength for signal Monte Carlo and a 10\% sample of the full one year dataset.}
  \label{simp_fig}
 \end{figure}

The challenge of separating a cascade signal from the overwhelming background of downgoing air-shower muons is significant.  In its 22 string configuration, $\sim$10 billion events triggered the IceCube detector in one year of operation.  Of these, only $\sim$10,000 are expected to be atmospheric neutrino-induced cascades.  Because the atmospheric $\nu_{\mu}$ and  $\nu_{e}$ fluxes differ \cite{Bartol-flux}, these  $\sim$10,000 events are unequally distributed among the different cascade signal classes.  For each  $\nu_{e}$, we expect $\sim$1.3  $\nu_{\mu}$ neutral-current events and $\sim$2.9  $\nu_{\mu}$ charged-current events where the hadronic cascade from the interaction vertex is inside the detector (so-called ``starting events'').        

To begin the analysis, a fast filter was developed to run online at the South Pole to select promising candidate events for satellite transmission to the northern hemisphere.  The filter selected events with a spherical topology that were not good fits to relativistically moving tracks.  After this online filter, each event was reconstructed according to track and cascade hypotheses using hit timing information, and well-reconstructed down-going tracks were thrown out.

A new, analytic energy reconstruction method for cascades was developed that takes into account the significant depth variation of the optical properties of the glacial ice at the South Pole~\cite{cascade-reco-paper}.  Several more topological variables with good separation power were also calculated for each event. 

The main background for neutrino-induced cascade searches comes from the stochastic energy losses suffered by cosmic ray muons as they pass through the ice surrounding the optical sensors.  Two basic variables are employed to reduce this background.  First, we measure how far inside the geometric volume of the detector the reconstructed cascade vertex lies.  Muons with a large stochastic energy loss far inside the detector are more likely to leave early hits in outer sensors and can thus be rejected.  Second, background separation becomes easier as the cascade energy increases.  This is because the more energetic stochastic losses that mimic neutrino-induced cascades would have to come from more energetic muons, which are more likely to leave additional light that will allow for their identification.  We therefore expect that more energetic cascades deep inside the detector will be the easiest signal to separate from background.

Along these lines, several neural networks were trained on 12 topological and reconstruction-based variables, including reconstructed energy and a measure of containment within the detector.  The product of these variables is taken as the final discriminating cut variable.  Figure 1 shows the number of remaining events as a function of the cut on this final variable for events that reconstruct above 5 TeV for signal Monte Carlo and a 10\% sub-sample of the available data.  

While nothing can yet be concluded from the 10\% data sample alone, the full dataset, which will be presented in this talk, may show signs of converging to the signal expectation.  Figure 2 shows the true neutrino energy at the earth's surface for the three classes of simulated signal.

 \begin{figure}[!t]
  \centering
  \includegraphics[width=2.5in]{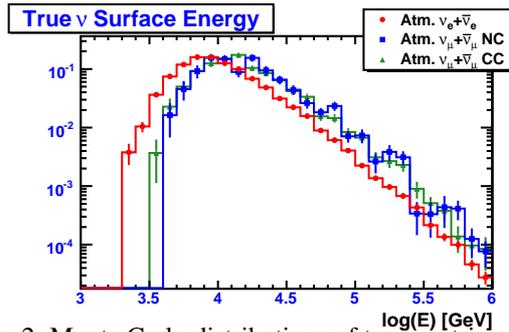}
   \caption{Monte Carlo distributions of true neutrino energy at the earth's surface for events surviving a final variable cut value of 0.73.}
  \label{simp_fig2}
 \end{figure}
 
 \bibliography{IEEEabrv,icrc1311}
 \bibliographystyle{IEEEtran}
  
\end{document}